\documentclass[aps,a4paper,superscriptaddress,twocolumn]{revtex4}
\usepackage{tensor}
\usepackage{graphicx}
\graphicspath{ {images/} }
\usepackage{amsmath}
\usepackage{amssymb}
\usepackage{enumerate}
\usepackage{subfigure}
\usepackage{tabularx}
\usepackage{lipsum} 
\usepackage{float} 
\usepackage[colorlinks=true, pdfstartview=FitV, linkcolor=blue, citecolor=red, urlcolor=black, breaklinks=true]{hyperref}
\newcommand{\be}{\begin{equation}}
\newcommand{\ee}{\end{equation}}
\newcommand{\ben}{\begin{eqnarray}}
\newcommand{\een}{\end{eqnarray}}
\newcommand{\bes}{\begin{subequations}}
	\newcommand{\ees}{\end{subequations}}
\def\bal#1\eal{\begin{align}#1\end{align}}

\newcommand{\LL}{{\mathcal L}}

\newcommand{\vfi}{\mathrm{\varphi}} 
 
\newcommand{\pu}{\mathrm{\partial_{\mu}}}

\newcommand{\Pu}{\mathrm{\partial^{\mu}}}

\newcommand{\pb}[1]{\ensuremath{\partial_{#1}}}
\newcommand{\lie}{$\rm{U(1)}\ $}

\begin{document}
	\title{Impurity-like solutions in vortex systems coupled to a neutral field}
	
	\author{D. Bazeia}\affiliation{Departamento de F\'\i sica, Universidade Federal da Para\'\i ba, 58051-970 Jo\~ao Pessoa, PB, Brazil}
	\author{M.A. Liao}\affiliation{Departamento de F\'\i sica, Universidade Federal da Para\'\i ba, 58051-970 Jo\~ao Pessoa, PB, Brazil}
	\author{M.A. Marques}\affiliation{Departamento de Biotecnologia, Universidade Federal da Para\'iba, 58051-900 Jo\~ao Pessoa, PB, Brazil}\affiliation{Departamento de F\'\i sica, Universidade Federal da Para\'\i ba, 58051-970 Jo\~ao Pessoa, PB, Brazil}
	
	\begin{abstract}
		In this work, a Maxwell-Higgs system is coupled to a neutral scalar field that engenders $Z_2$ symmetry. At critical coupling, the resulting field equations may be identified with those of a Maxwell-Higgs model doped with an impurity whose form changes according to properties of the neutral scalar field, such as the topological sector and position of its zeros. This allows for an interpretation of parameters appearing in impurity models in terms of properties of the kink-like defect, and provides a convenient way to understand and generate impurities. By solving the first order equations, we found vortices with a novel internal structure in relation to standard solutions. The procedure was also adapted to generate impurities for Chern-Simons-Higgs theory.
	\end{abstract}
	
	
	\date{\today}
	\maketitle

An extension of the three dimensional Maxwell-Higgs model meant to allow for the presence of magnetic impurities has been considered in Refs.~\cite{Hook, Tong}. Maxwell-Higgs theory has long been known to engender vortex solutions~\cite{Nielsen}, and is closely associated with superconductivity, as its static equations are those of Ginzburg-Landau theory, which have been used to predict the existence of magnetic vortex lines in type-II superconducting materials~\cite{abrikosov}. The impurity may be added through the inclusion of an additive term of the form $B\sigma(\chi)$ and a corresponding deformation in the potential. Bogomol'nyi's procedure~\cite{bogo} may be adapted to this model, giving rise to minimal energy solutions at critical coupling. Doping with impurities may be used to change the magnetic properties of superconducting materials, and has interesting consequences for vortex motion and scattering~\cite{Cockburn, Ashcroft}. Impurities have also found applications in holography~\cite{Benincasa,BenincasaII,Hashimoto,Erdmenger,Kachru}.  It was shown in~\cite{Hook} that such an extension preserves half of the supercharges in supersymmetric QED, and may be used to study the interaction of a strongly coupled system with a lattice of impurities. In this paper, it is revealed that, under certain assumptions, these impurities may be interpreted in terms of a kink-like defect coupled to a Maxwell-Higgs system. We show that some properties of the impurity may be understood in terms of physically meaningful parameters of the kink.

Consider a \lie gauge theory in which a complex-valued scalar field $\varphi$ is coupled to a gauge field $A_{\mu}$ in three spacetime dimensions, with metric tensor $\eta_{\mu\nu}= \rm{diag}(1,-1,-1)$, in natural units. These fields are now coupled to a neutral scalar field $\chi$, which will be used to generate the impurities in this work. Coupling between a Maxwell-Higgs system and a real scalar field has been previously investigated in Refs.~\cite{internal, multilayered}, where the neutral scalar field was used to generate internal structure in vortex profiles in models with generalized permeability. In the present case, the field equations are obtained from the Lagrangian density
\begin{equation}\label{Lag}
\begin{split}
\LL=&-\frac{1}{4}F_{\mu\nu}F^{\mu\nu} + D^{\mu}\vfi\overline{D_{\mu}\vfi} - \frac{\lambda}{2}(1 + \sigma(\chi)-|\vfi|^2 )^2\\
 &+ B\sigma(\chi) +\frac{1}{2}\pu\chi\Pu\chi - \widetilde{V}(\chi, \mathbf{x}),
\end{split}
	\end{equation}
where  $D_{\mu}=\pb{\mu} + ieA_{\mu}$, $F_{\mu\nu}=\pb{\mu} A_{\nu} -\pb{\nu} A_{\mu}$ and $B=F_{21}$. The function $\widetilde{V}(\chi,\mathbf{x})$ is such that its minimum value is attained for $\chi=v_{\pm}$, with $v_{-}\neq v_{+}$, thus allowing for the emergence of a topological defect. In this work, $\sigma(\chi)$ is taken as a bounded, differentiable function, and we are mostly interested in choices that go to zero when $\chi$ approaches one of its vacuum values. For a topological solution $\chi$, the last assumption implies an impurity that is localized, since such a defect is appreciably different from a vacuum field only at a finite region of the plane. With these assumptions, vortex solutions must satisfy $\varphi\to e^{ie\Lambda(t,\mathbf{x})}$, $\mathbf{A}=-\nabla \Lambda(t,\mathbf{x})$ asymptotically to prevent infinite energy. These conditions lead to a quantized magnetic flux $\Phi=2\pi n/e$, where $n\in\mathbb{Z}$ is a topological winding number.

The coupling constant $e$ will henceforth be taken as unity, and we note that it may be easily reintroduced in the Lagrangian. We shall also specialize in time-independent solutions at critical coupling, defined by $\lambda=1$ in our units. The equations of motion obtained from~\eqref{Lag} under these assumptions are
\begin{subequations}\label{SO}
\begin{align}
&\nabla^2\chi=-\sigma_{\chi}\left[B - (1 + \sigma(\chi) -|\varphi|^2)\right] + \widetilde{V}_{\chi}, \label{SO1}\\
&D_{k}D_{k}\varphi=(1 + \sigma(\chi) -|\varphi|^2)\varphi, \label{SO2}\\
&\epsilon_{jk}\pb{j}(B - \sigma(\chi))=i(\overline{\varphi}D_k\varphi - \varphi\overline{D_k\varphi}), \label{SO3}  %+ \frac{W_{\chi}W_{\chi\chi}}{r},
\end{align} 
\end{subequations}
where $\widetilde{V}_{\chi}\equiv \partial\widetilde{V}/\partial\chi$. The remaining equation, found from variation of the action with respect to $A_0$, is Gauss's law, which is trivially satisfied by $A_{0}=\pb{t}\varphi=0$, as long as the electric field is absent. Note that, as $\sigma(\chi)\to 0$, Eqs.~\eqref{SO2} and~\eqref{SO3} are deformed into the usual Ginzburg-Landau equations. 

One would like to solve~\eqref{SO1} independently of $\varphi$ and $\mathbf{A}$ in order to substitute $\chi$ as a function of the coordinates in Eqs.~\eqref{SO2} and~\eqref{SO3}. This is however not possible in general, because of the first term in the right-hand side of Eq.~\eqref{SO1}, which makes it impossible, in general, to solve this equation without knowledge of the magnetic field. However, we note that this term vanishes  when the self-dual vortex equations
	\begin{subequations}\label{BPS}
		\noindent\begin{minipage}{.5\linewidth}
			\begin{equation}
			B=(1 + \sigma -|\vfi|^2), 
			\end{equation}
			\vspace{0.5pt}
		\end{minipage}%
		\begin{minipage}{.5\linewidth}
			\begin{equation}
			(D_{1} + iD_2)\varphi=0, 
			\end{equation}
			\vspace{0.5pt}
		\end{minipage}
	\end{subequations}
are solved. Moreover, the energy functional of static solutions may be written in the form $E=E_1 + E_2$, where

\begin{equation}\label{E_1}
\begin{split}
E_1	&=\int d^2x \left[\frac{B^2}{2} -\sigma(\chi)B + |D_1\varphi|^2 + |D_2\varphi|^2 \right.\\
& \ \  \ \ \ \ \ \ \ \ \ \ \ \   \bigg.  + \frac{1}{2}(1 + \sigma(\chi) -|\varphi|^2)^2 \bigg] ,
\end{split}
\end{equation}
and 
\begin{equation}\label{E_2}
E_2	=\int d^2x \left(\frac{1}{2}\nabla\chi\cdot\nabla\chi + 	\widetilde{V}(\chi, \mathbf{x}) \right).
\end{equation}
Through a Bogomol'nyi procedure~\cite{bogo}, one may verify that $E_1$ is minimized to the value $E_1=2\pi n$ by solutions of~\eqref{BPS}. In that case, the self-dual equations imply $\nabla^2\chi=\widetilde{V}_{\chi}$, which is precisely the second order equation one would obtain from the condition $\delta E_2=0$. It thus suffices to find a stable solution $\chi(x,y)$ which is a stationary point of $E_2$ in order to generate an impurity. This may be achieved through the choice
\begin{equation}\label{pot}
	\widetilde{V}(\chi, r)=\frac{W_{\chi}^2}{2r^2},
\end{equation}
where $W_{\chi}$ is the derivative of an auxiliary function $W(\chi)$. A self interaction of this form may arise naturally in an effective description of a scalar field interacting with other field(s), see~\cite{effective} for example. In particular, the potential~\eqref{pot} was introduced in \cite{stable} as a way of evading Derrick's argument~\cite{Derrick} which, in canonical models, forbids the existence of stable static kink-like solutions in flat spacetime with more than one spatial dimension. Such a solution is found if one takes $\chi$ as a radially symmetric field~\cite{stable}. By completing squares we note that these assumptions give rise to the bound $E_2\geq 2\pi|W(r\to\infty) - W(r=0)|$. The energy saturates this bound, being thus minimized, if and only if $\chi$ solves either of the first order equations 
\begin{equation}\label{FOchi}
\chi'(r)=\pm\frac{W_{\chi}}{r},
\end{equation}
where the prime denotes derivation with respect to the radial coordinate. The positive and negative signs in the above equation are related by the transformation $r\to 1/r$, so we shall henceforth consider only upper sign in~\eqref{FOchi}. Together with~\eqref{BPS}, this equation gives rise to a system of first order equations whose solutions satisfy \eqref{SO}. Eq.~\eqref{FOchi} may be solved independently of $\varphi$ and $\mathbf{A}$, and the result substituted in~\eqref{BPS} to recover the first order equations found in~\cite{Tong}. However, our model differs from those usually considered in some important ways, most notably in the fact that the impurity is not a fixed function that is added to the Lagrangian; it changes according to the boundary conditions the neutral scalar field is subject to. A non-topological $\chi$ would be expected to decay into the vacuum, thereby resulting in $E_2=0$ and $E_1$ identical to the static energy functional of impurity-free Maxwell-Higgs theory. 

Because radial symmetry was assumed for $\chi$, the impurity thus obtained is a function of the radial coordinate alone, which corresponds to the form often considered in investigations of vortex-impurity solutions, see~\cite{Cockburn,Ashcroft} for example.  However, any static stable stationary point of $E_2$ may in principle be fed into $\sigma(\chi)$, so impurities devoid of axial symmetry may be generated by this method. Since $\widetilde{V}$  does not appear in $E_1$, the argument is essentially unchanged if one makes a different choice of this function, as long as stable solution is found. However, defects of this kind are not easy to find, so we shall assume~\eqref{FOchi} holds for the remainder of this paper. 

At low energies, the dynamics of topological defects may often be approximated as motion in the Moduli space~\cite{geodesic, Ruback, geodesicII, Fuertes}. Under this approximation, the vortices move through geodesics in the space generated by solutions of the Bogomol'nyi equations, with a metric induced by their kinetic energy. In the Maxwell-Higgs model, the validity of this approximation has been rigorously proved~\cite{Stuart}. Since Eqs.~\eqref{BPS} are satisfied throughout the motion, solutions of~\eqref{FOchi} will still solve~\eqref{SO1}, so this equation may still be used as an approximation in this dynamical scenario. This reasoning works as long as the neutral field may be considered fixed to good approximation, meaning that the scattering of the kink by the vortices can be neglected. In that case, $\chi$ generates a background in which the vortices move, in a way that is not unlike the argument given by Tong and Wong, who showed that impurities may be seen as frozen, heavy vortices in a $\rm{U(1)}\times\rm{U(1)}$ theory~\cite{Tong}. If the kink is also moving, the first order equations may still be useful as long as the geodesic approximation holds and $\sigma(\chi)$ changes sufficiently slowly as a function of t. In that case, the neutral scalar field solves $\Box\chi + \widetilde{V}_{\chi}=0$, and the result generates, at a time $t$ close to an instant $t_0$, an impurity $\sigma=\sigma(\mathbf{x},t_0)$. At different times $t_0$, the solution looks like a Maxwell-Higgs system modified by  different impurity functions.

As in the Maxwell-Higgs model, one may use the definition $\xi\equiv \ln|\varphi|$~\cite{Taubes} to condense Eqs.~\eqref{BPS} into the equality
	\begin{equation}\label{xidelta}
	\nabla^2\xi + 1 -\exp(2\xi)=2\pi\sum_{j}\delta(\mathbf{x} - \mathbf{x_j}) -\sigma(\chi(r)) ,
	\end{equation} 
where $\mathbf{x_j}$ are the zeros of $\varphi$, counted as many times as their multiplicity. We see that the impurity provides an additional source in Eq.~\eqref{xidelta}, as is of course the case in the original model~\cite{Cockburn}. Here, however, the location of this source is affected by the position of the kink-like defect. An important example occurs when
\begin{equation}\label{delta}
\sigma(\chi)=-2\pi m\delta(\chi),
\end{equation}
where $m$ is a positive integer. Delta impurities have been examined in~\cite{Cockburn}, where it was shown that, although a Lagrangian description cannot give rise to an impurity of this form, the equations of motion remain well defined when they are used. Moreover, delta impurities may be obtained as a limiting case for Gaussian impurities. 

With that choice,~\eqref{xidelta} takes the form of a Taubes equation with $n+m$ delta sources, as if $m$ extra vortices had been added to the system. Unlike the $n$ sources identified with zeros of $\varphi$, which may take arbitrary positions, the impurity~\eqref{delta} always behaves as $m$ vortices superimposed at the point where $\chi=0$, and which cannot be separated. Since kink solutions only connect two minima of the potential, such a zero must be unique, as Eq.~\eqref{FOchi} prevents $\chi'$ from changing its sign before $\chi$ reaches a minimum. If $\chi(\infty)=-\chi(0)= v$ are taken as boundary conditions, the position of the kink is not constrained by~\eqref{FOchi}. As an example, let $W_{\chi}=1-\chi^2$. This choice leads to the $\chi^4$ potential, which engenders $Z_2$ symmetry in the theory and has the kink-like solution~\cite{stable}
\begin{equation}\label{chi4}
\chi=\frac{r^2-r_0^2}{r^2 + r_0^2}.
\end{equation}
Here, $r_0$, which defines the zero of $\chi$, arises as a constant of integration, and may take an arbitrary value. For the sigma function~\eqref{delta}, this implies that the location of the $m$ delta sources introduced by the impurity in~\eqref{xidelta} is fixed by this integration constant, and may take any value. This is an important feature of the model, as it implies that not only there are multiple possible impurity functions - which lead to different magnetic properties for the $\rm{U(1)}$ subsystem - for a given sigma function, but also that these impurities may change when the position of the kink is changed. 

Investigation of the field equations is greatly simplified when rotational symmetry is assumed, as first done by Abrikosov~\cite{abrikosov}. In the static case, we assume $A_0=F_{0k}=0$ and look for solutions of the form
\begin{subequations}\label{Ansatz}
	\noindent\begin{minipage}{.5\linewidth}
		\begin{equation}\label{g}
		\varphi=g(r) e^{in\theta},
		\end{equation}
		\vspace{0.5pt}
	\end{minipage}%
	\begin{minipage}{.5\linewidth}
		\begin{equation}\label{a}
		\vec{A}=\frac{n-a(r)}{r}\hat{\theta},
		\end{equation}
		\vspace{0.5pt}
	\end{minipage}
\end{subequations}
subject to the boundary conditions $g(0)=a(\infty)=0$, $a(0)=n$, $g(\infty)=1$. When~\eqref{FOchi} is solved, $\sigma$ may be substituted as function of $r$, so that Eqs.~\eqref{BPS} become the pair of ordinary differential equations

\begin{subequations}\label{FO}
	\noindent\begin{minipage}{.5\linewidth}
		\begin{equation}\label{bps1}
		g' =\frac{ag}{r}, 
		\end{equation}
		\vspace{0.5pt}
	\end{minipage}%
	\begin{minipage}{.5\linewidth}
		\begin{equation}\label{bps2}
		-\frac{a'}{r}=1 + \sigma(r) -g^2 .
		\end{equation}
		\vspace{0.5pt}
	\end{minipage}
\end{subequations}
The integrands in~\eqref{E_1} and~\eqref{E_2} give rise to an energy density $\rho=\rho_1 + \rho_2$. When the first order equations are solved, they may be written as 

 \begin{subequations}
 	\noindent\begin{minipage}{.5\linewidth}
 		\begin{equation}\label{rho1}
	 	\rho_1= B^2 + 2g'^2 - B\sigma
 		\end{equation}
 		\vspace{0.5pt}
 	\end{minipage}%
 	\begin{minipage}{.5\linewidth} 
 		\begin{equation}\label{rho2}
 		\text{and}\hspace{20pt}\rho_2= \chi'^2 ,
 		\end{equation}
 		\vspace{0.5pt}
 	\end{minipage}
 \end{subequations}
where $B=-a'/r$. $\rho_1$ is formally identical to the energy density of a system composed of a symmetric vortex and an impurity $\sigma(\chi(r))$. As was the case in the original model, $\rho_1$ may not be everywhere positive for some choices of $\sigma$. However, in our model there is another contribution $\rho_2$ that is strictly positive for any nontrivial solution of~\eqref{FOchi}. This allows for an everywhere positive energy density in many cases, even when $\rho_1 < 0$ for some $r$. This feature, found in all solutions presented in this work, provides a physical interpretation for the negative sign found in the energy density of vortex-impurity systems, which may be seen as a consequence of the fact that the energy of another subsystem is not taken into account when $\rho_1$ is considered in isolation. 

\begin{figure}
	\centering
	\includegraphics[width=4.2cm]{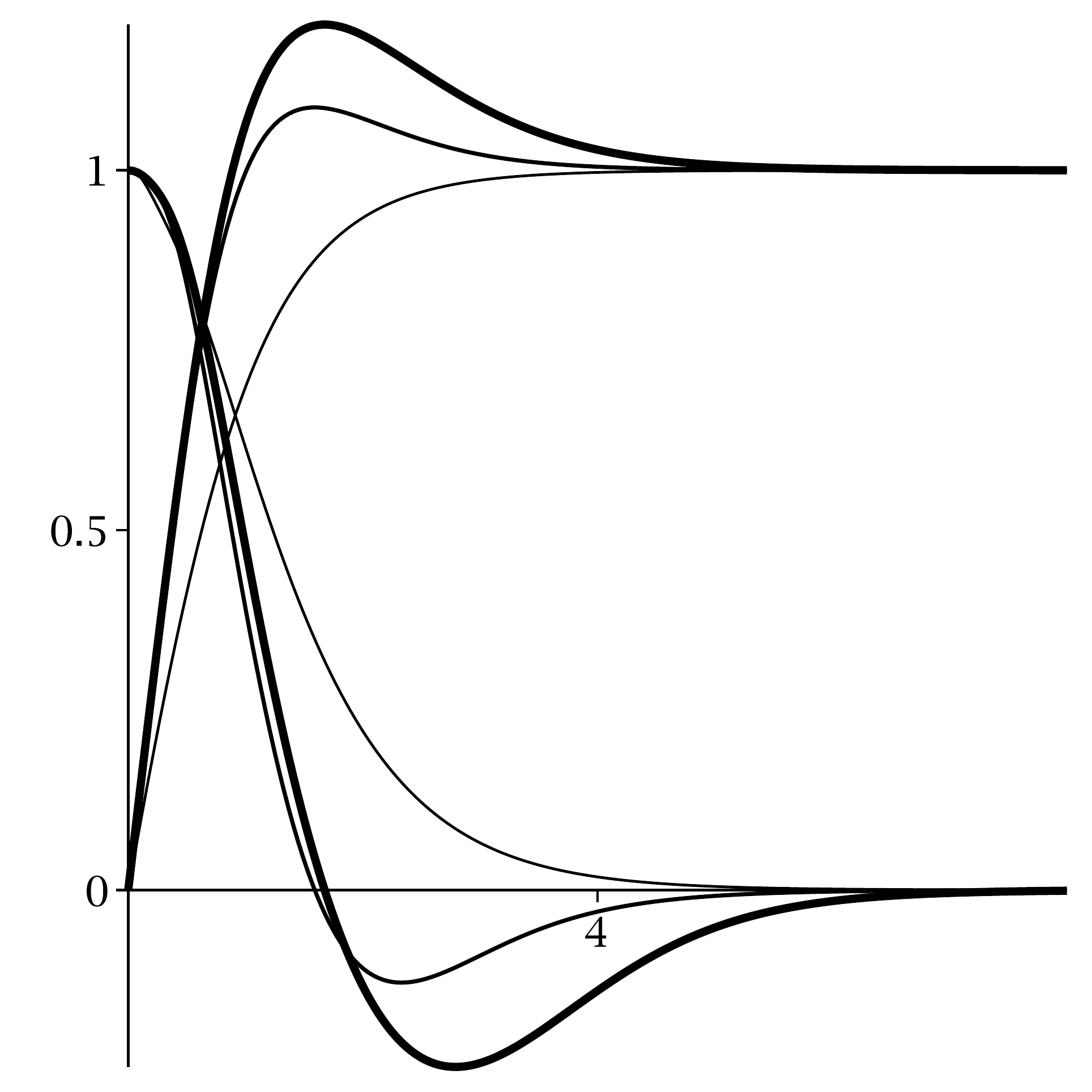}
	\includegraphics[width=4.2cm]{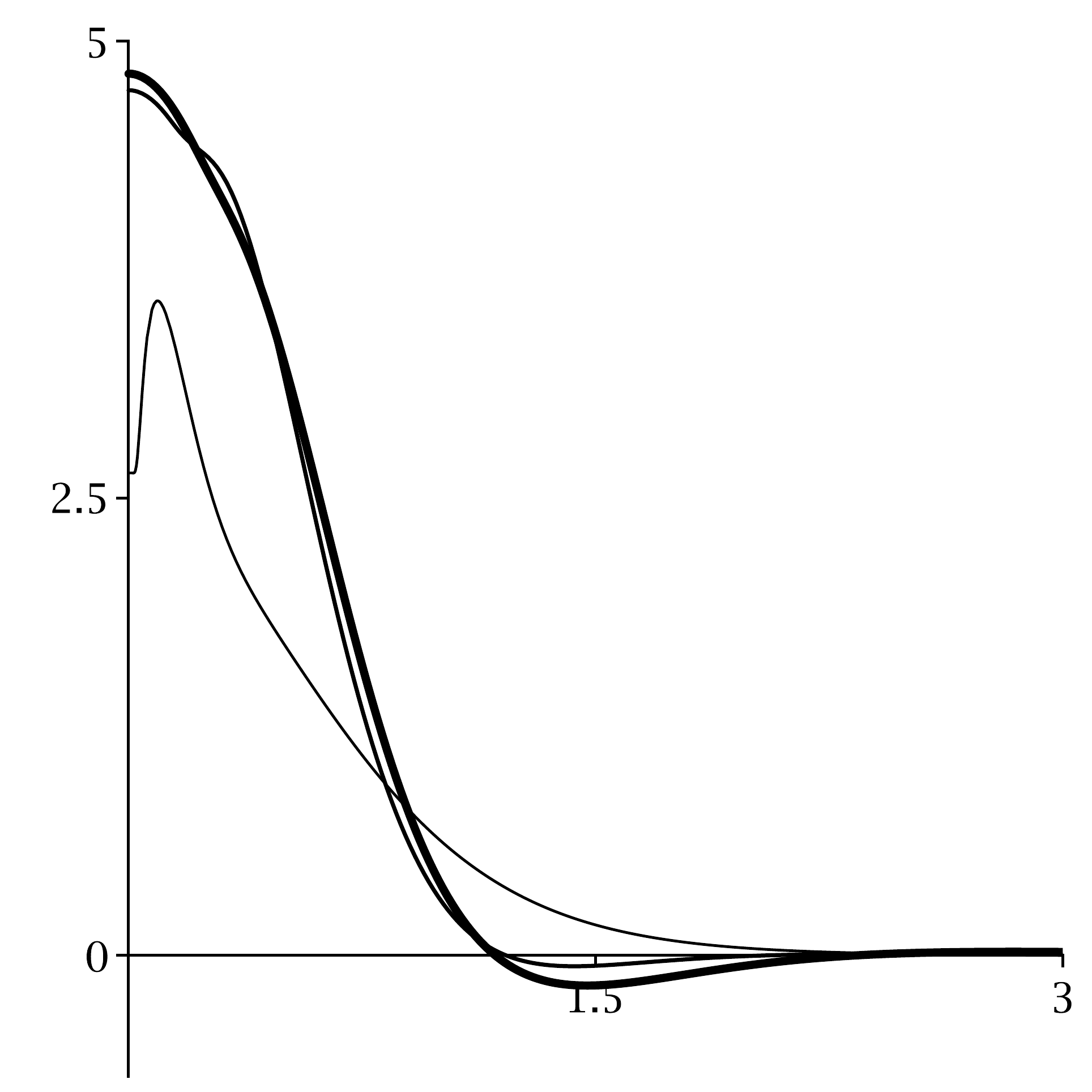}
	\caption{Solutions $g(r)$ (ascending lines) and $a(r)$ (descending lines) of Eq.~\eqref{FO} (left) and the associated energy density $\rho_1$ (right) for $\sigma(\chi)$ satisfying~\eqref{sigma} with $\alpha=1$, and generated by a neutral scalar field of the form~\eqref{chi4}. Here, $n=1$, and line width increases with $r_0^2$, which takes the values $0.01$, $0.5$ and $1$.}
	\label{fig1}
\end{figure}

We may exemplify our results with a family of models with a $\sigma$ function of the form
\begin{equation}\label{sigma}
\sigma(\chi)=\alpha e^{-\frac{\chi^2}{1-\chi^2}},
\end{equation}
for $\chi^2\neq 1$ and $\sigma(\chi)=0$ for $\chi^2=1$. Here, $\alpha$ is a parameter whose magnitude controls the maximum height of $\sigma$. This function falls rapidly to zero when $\chi\to \pm 1$. If the neutral scalar field is taken in the form~\eqref{chi4}, there results a bell-shaped impurity, which is zero both at the origin and asymptotically, and has a maximum at $r=r_0$ for positive $\alpha$. Thus, the zero of the kink controls the position of the peak of $\sigma(r)$, which moves to the right as $r_0$ is increased. In Fig.~\ref{fig1}, we depict the solutions of~\eqref{FO} for some choices of $r_0$ in the case $\alpha=1$, along with the corresponding energy density $\rho_1$.  We see that, unlike the impurity-free version of the theory, $g(r)$ and $a(r)$ are not monotone functions and instead have, respectively, a maximum and a minimum occurring for a finite value of $r$. The height/depth of this maximum/minimum is changed according to the relative distance between vortex and kink. In particular, for a very small $r_0$ the vortex profiles look increasingly like their impurity-free counterparts, indicating that, in the limit where $r_0\to 0$, which represents a situation where the zeros of the kink and the vortex coincide, $(g, a)$ would look like a standard solution. Despite the change effected by different choices of $r_0$, the position of this point, and therefore of the peak of $\sigma$, do not change the total energy of the configuration.

\begin{figure}[h]
	\centering
	\includegraphics[width=4.8cm]{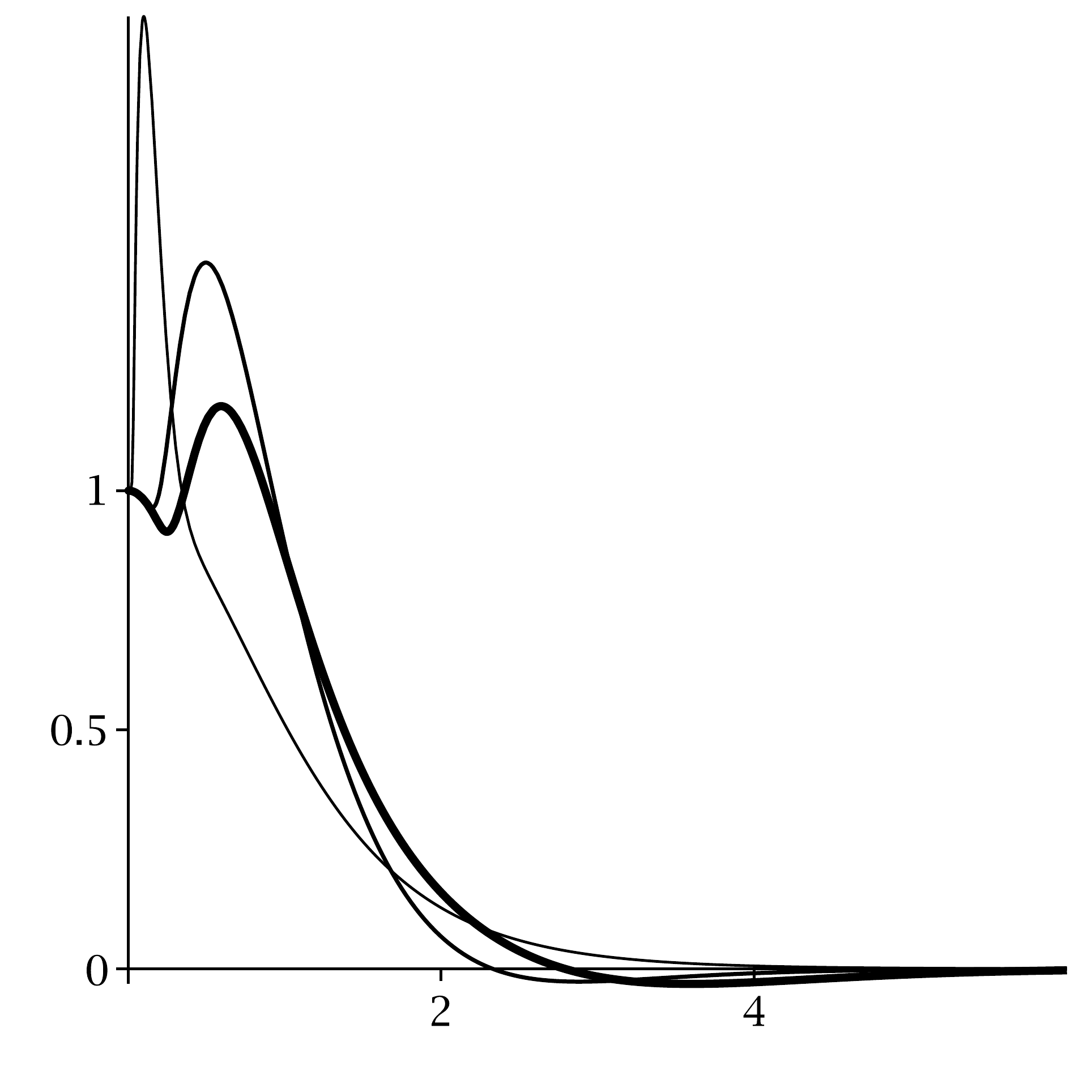}
	\caption{Magnetic field $B(r)$ relative to a $n=1$ solution of~\eqref{FO} for an impurity of the form~\eqref{sigma}, with $\alpha=1$, and generated by the kink-like solution~\eqref{chi4}. Here, line width increases with $r_0^2$, which takes the values $0.01$, $0.5$ and $1$. }
	\label{fig2}
\end{figure}

In Fig.~\ref{fig2}, we depict the magnetic field for the same system. At $r=0$, $\sigma(r)$ vanishes, and at this point the magnetic field equals one in all cases, the same value that one would find in the impurity-free theory with this choice of units. At a neighborhood of the vortex center, $\sigma(r)$ increases at a rate controlled by the zero of $\chi$. In such a neighborhood, $B(r)$ may be either increasing, as seen in Fig.~\ref{fig2} for $r_0=0.1$, or decreasing, as happens for the other choices of $r_0$ depicted in the same figure. As $r_0$ increases, the peak of $B(r)$ is moved to the right. Also noteworthy is the flux inversion caused by a change of sign in the magnetic field, which occurs at different positions for the profiles depicted in Fig.~\ref{fig2}.

\begin{figure}[h]
	\centering
	\includegraphics[width=4.2cm]{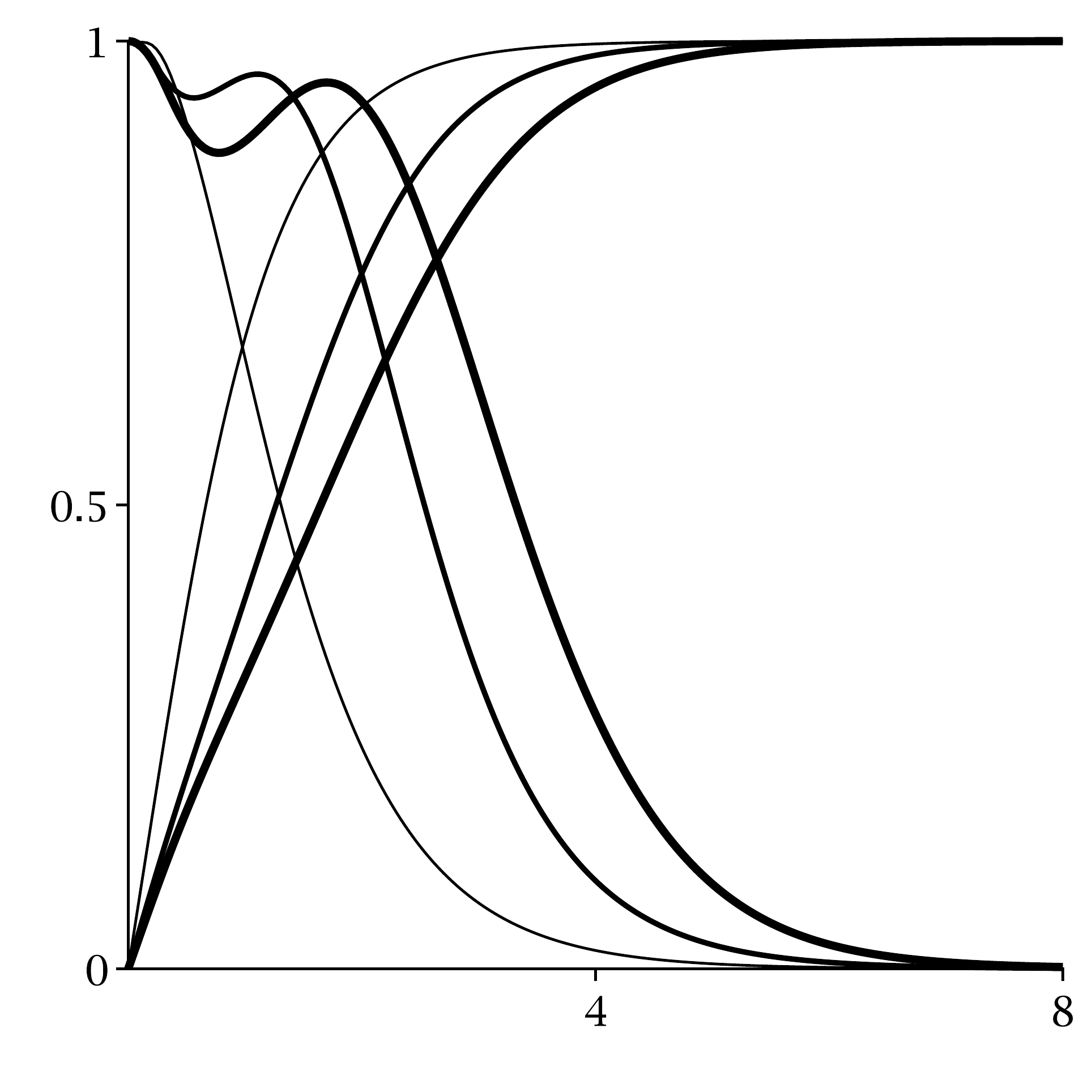}
	\includegraphics[width=4.2cm]{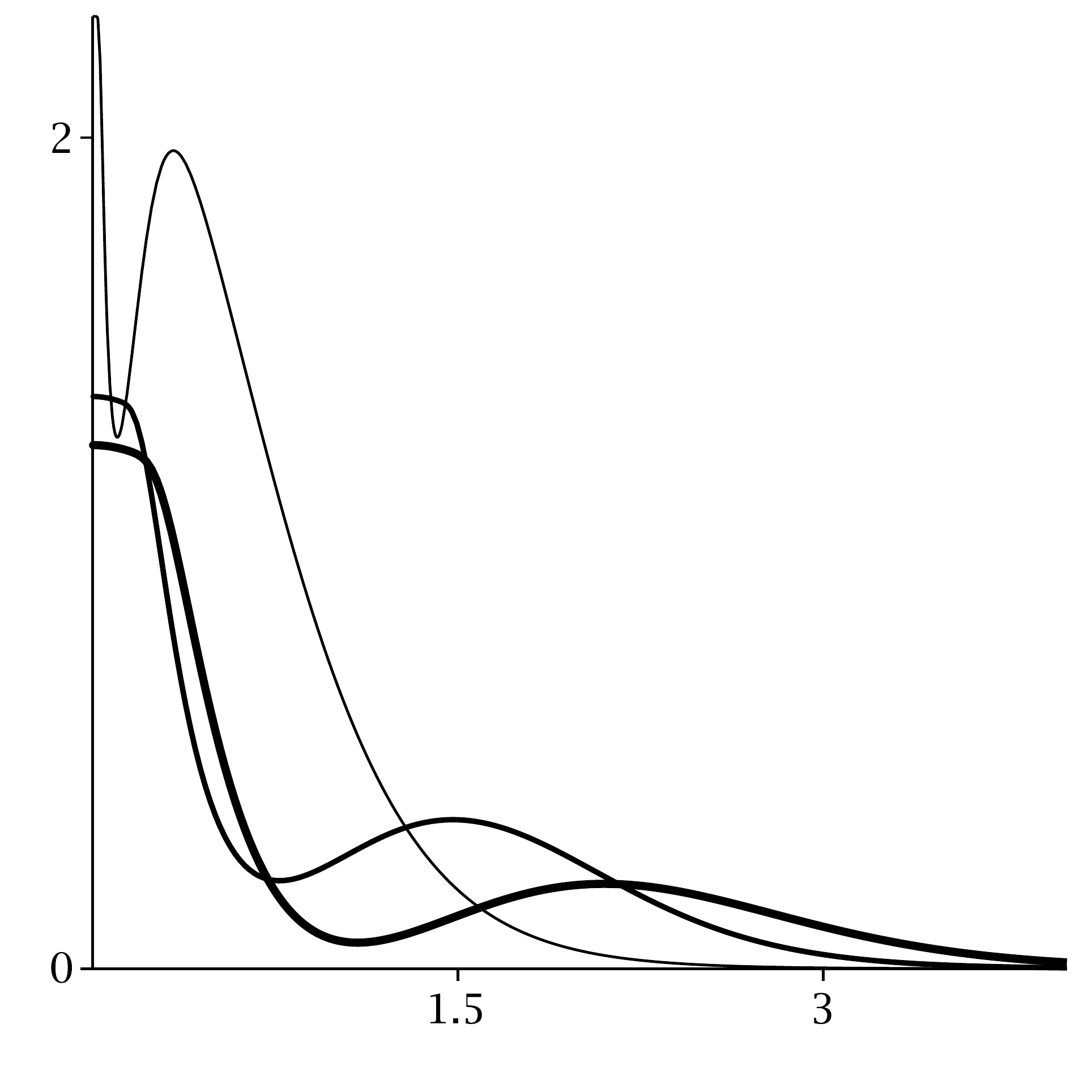}
	\caption{Solutions $g(r)$ (ascending lines) and $a(r)$ (descending lines) of Eq.~\eqref{FO} (left) and the associated energy density $\rho_1$ (right) for $\sigma(\chi)$ satisfying~\eqref{sigma} with $\alpha=-1$, and generated by a neutral scalar field of the form~\eqref{chi4}. Here, $n=1$, and line width increases with $r_0^2$, which takes the values $0.01$, $0.5$ and $1$.}
	\label{fig3}
\end{figure}

\begin{figure}[h]
	\centering
	\includegraphics[width=4.8cm]{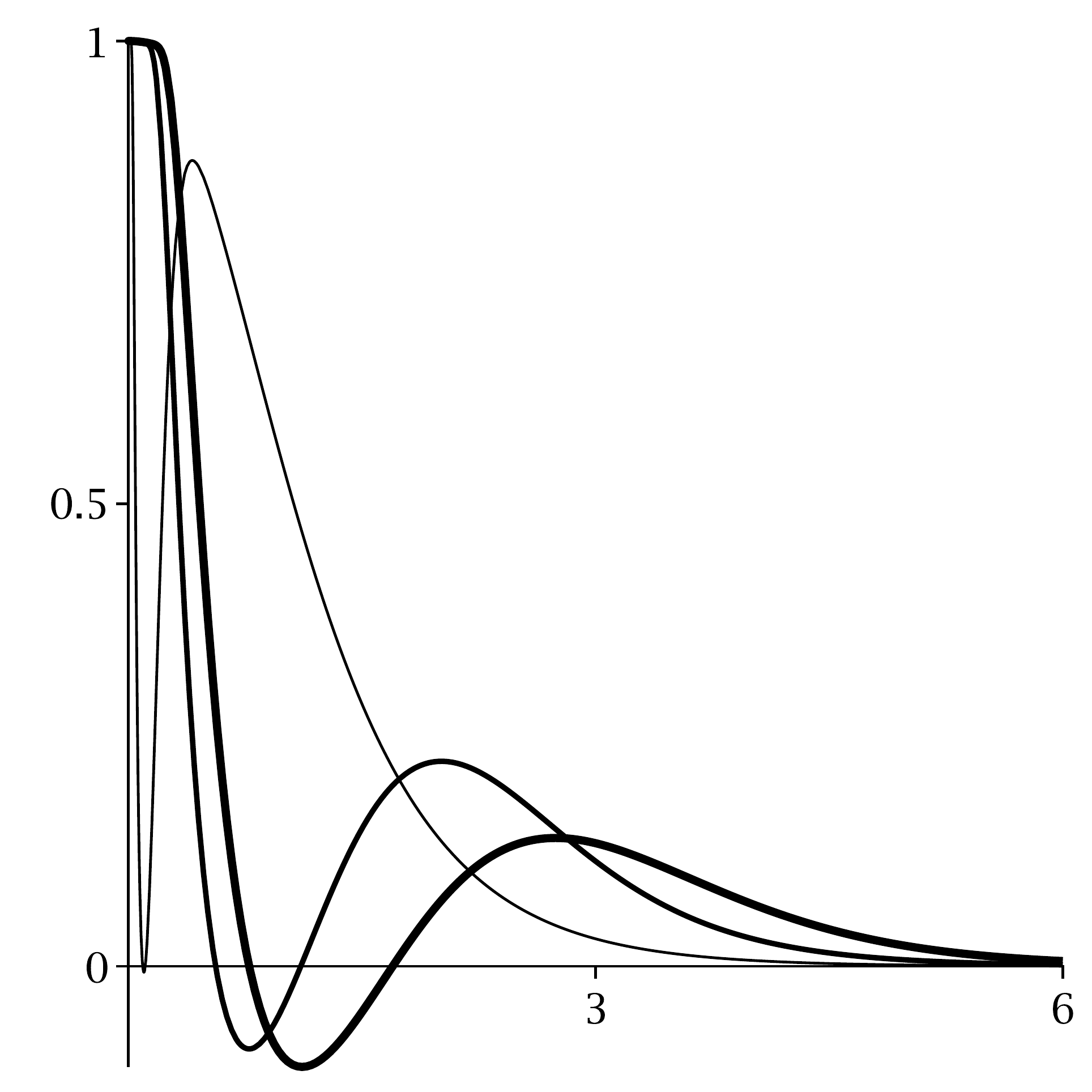}
	\caption{Magnetic field $B(r)$ relative to a $n=1$ solution of~\eqref{FO} for an impurity of the form~\eqref{sigma}, with $\alpha=-1$, and generated by the kink-like solution~\eqref{chi4}. Line width increases with $r_0^2$, which takes the values $0.01$, $0.5$ and $1$.  }
	\label{fig4}
\end{figure}

When $\alpha$ is negative, a very different behavior is found, as both $a(r)$ and $g(r)$ become strictly positive functions for all values of $r_0$ we have considered. As can be seen in Fig.~\ref{fig3}, $g(r)$ is a monotonically increasing function of $r$, a behavior which mirrors that of the standard Nielsen-Olesen solution, found in the impurity-free Maxwell-Higgs theory. On the other hand, $a(r)$ possesses internal structure that is not found in the standard model, having gained two local extrema, whose positions and magnitudes depend on $r_0$. Also noteworthy is the energy density of the vortex-impurity system, depicted in the same figure. Unlike the $\alpha=1$ case, we see that when $\alpha=-1$, $\rho_1(r)$ becomes a positive definite function for the three choices $r_0$ shown in the figure. Finally, we depict, in Fig.~\ref{fig4}, the magnetic fields associated to these solutions. We note the presence of a finite region, appearing as valley in the plot of $B(r)$, where the magnetic field changes its sign. The width and depth of this valley is seen to increase with $r_0$.


\begin{figure}[h]
	\centering
	\includegraphics[width=4.2cm]{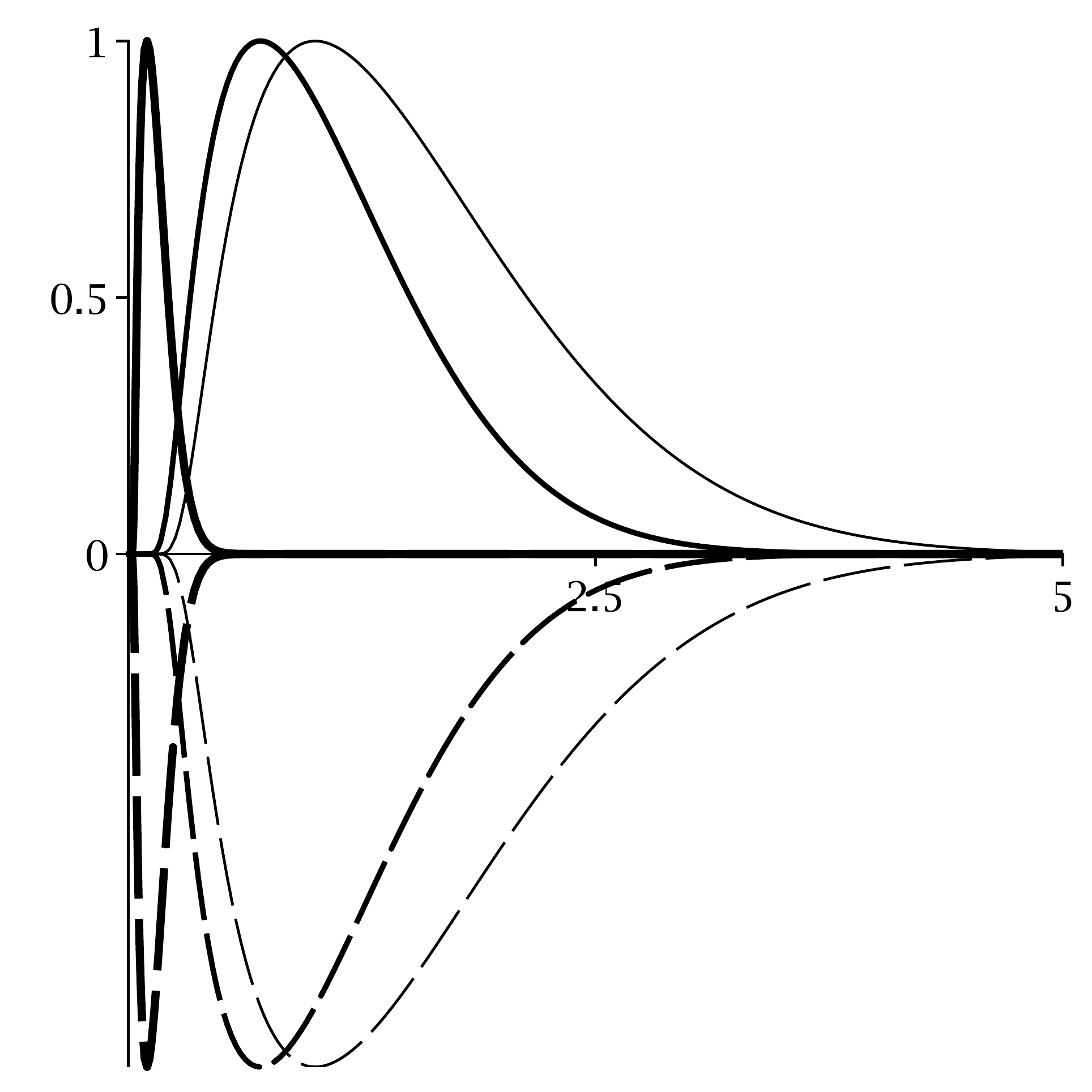}
	\includegraphics[width=4.2cm]{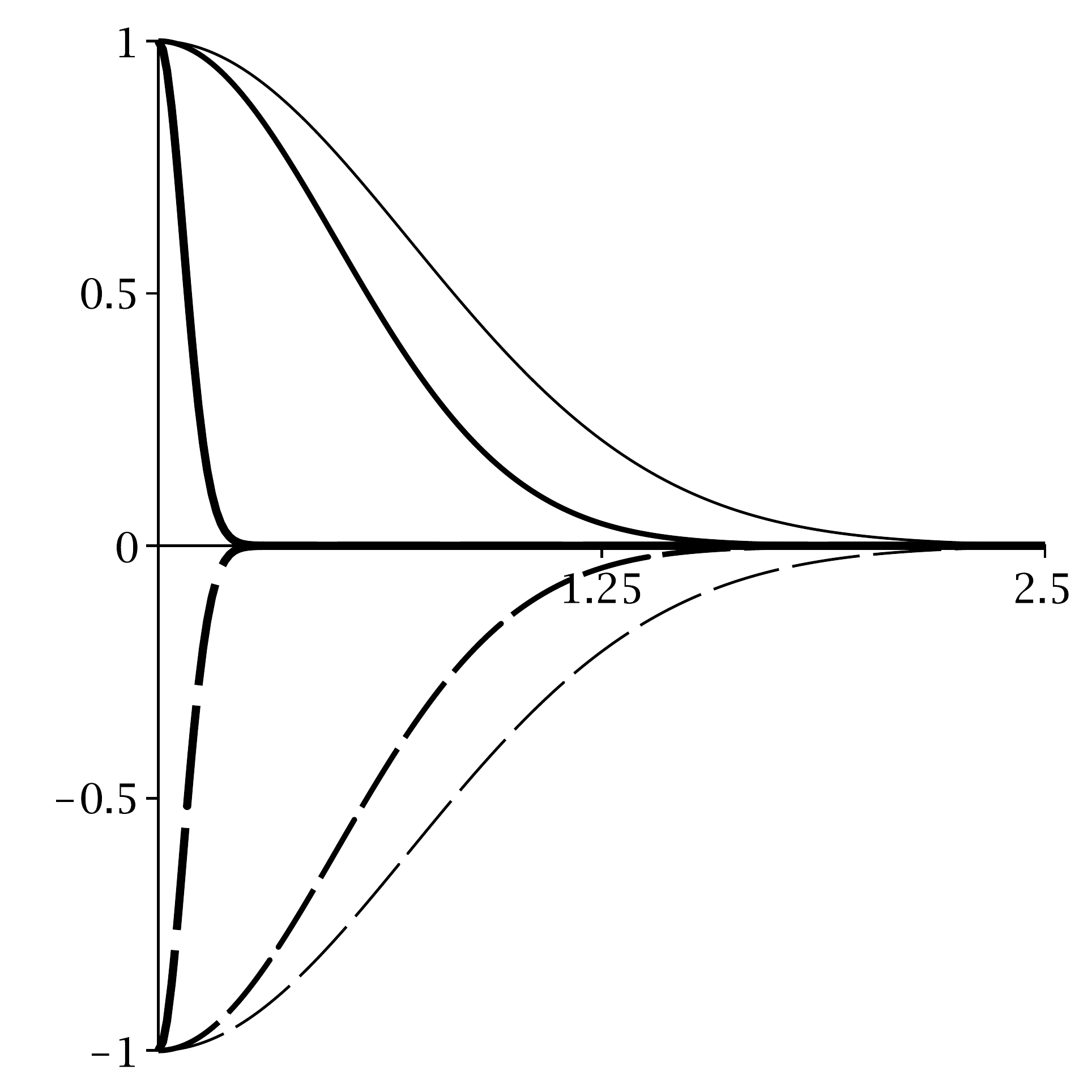}
	\caption{Impurity~\eqref{sigma} for $\alpha=1$ (solid lines) and $\alpha=-1$ (dashed lines), with a neutral scalar field given by~\eqref{chi4} (left) and~\eqref{chi6} (right). Line width increases with $r_0^2$, which takes the values $0.01$, $0.5$ and $1$.}
	\label{fig5}
\end{figure}

The same function $\sigma(\chi)$ gives rise to a very different impurity model when other choices of the auxiliary function $W(\chi)$ are considered. Indeed, let $W_{\chi}=\chi(1-\chi^2)$, which leads to a $\chi^6$ potential. In that case,~\eqref{FOchi} is solved by
\begin{equation}\label{chi6}
\chi=\frac{r}{\sqrt{r_0^2 + r^2}},
\end{equation}
where $r_0$ is again an integration constant, although it cannot be identified with the zero of $\chi$ anymore, which is always located at the origin in this example. In that case, the resulting impurity is $\sigma(r)=\alpha e^{-r^2/r_{0}^2}$, which, in contrast to the $\sigma(r)$ studied in the previous examples, is a monotonic function, with a maximum (if $\alpha>0$) or a minimum (if $\alpha<0$) at the origin. This impurity is of the form $\sigma(r)=\alpha e^{-\beta r^2}$ considered in~\cite{Cockburn, Ashcroft}, where the vortex solutions are shown. In our model, the constant $\beta=1/r_0^2$, which appears as a free parameter in those models, has a new interpretation, as it is specified by $r_0$, a characteristic parameter of the kink solution. The impurities generated by both types of kink-like sources are depicted in Fig.~\ref{fig5}. There is another important difference between the models generated by the $\chi^4$ and $\chi^6$ self-interactions, as the former leads to the same impurity regardless of the sign in~\eqref{FOchi}, while the latter does not. Indeed, the antikink solution obtained by the transformation $r\to1/r$ in~\eqref{chi6} leads to a $\sigma(r)$ which is zero at the origin and one at infinity. This behavior is opposite to the one displayed in Fig.~\ref{fig5}, and is owed to the fact that $|\chi(0)|\neq|\chi(\infty)|$ for this model. We thus realize that the topological sector of the neutral scalar field, which distinguishes a kink from an antikink, may also play a role in the profile of the impurity, even when only topological solutions are considered.

Although we have worked with the Abelian Maxwell-Higgs model, the method developed in this paper may be extended to generate impurities for other systems. An important variation is achieved by including an additive term $B\sigma(\chi)$ to the Chern-Simons-Higgs system considered in Refs.~\cite{Jackiw, Hong}. In that case, the equations of motion may be obtained from the Lagrangian
\begin{equation}\label{LagCS}
\begin{split}
\LL=&\frac{\kappa}{4}\epsilon^{\alpha\beta\gamma}A_{\alpha}F_{\beta\gamma} + D^{\mu}\vfi\overline{D_{\mu}\vfi} -\frac{|\varphi|^2(1+\sigma(\chi)-|\varphi|^2)^2}{\kappa}\\
&+ B\sigma(\chi) +\frac{1}{2}\pu\chi\Pu\chi - \widetilde{V}(\chi, \mathbf{x}),
\end{split}
\end{equation}
where $\kappa$ is a constant.  The first order equations are found by minimizing $E_2$, left unchanged, and $E_1$, now taken as
\begin{equation}\label{E_1_CS}
\begin{split}
E_1	&=\int d^2x \left[\frac{\kappa^2B^2}{4|\varphi|^2} -\sigma(\chi)B + |D_1\varphi|^2 + |D_2\varphi|^2 \right.\\
& \ \  \ \ \ \ \ \ \ \ \ \ \ \   \bigg.  + \frac{|\varphi|^2}{\kappa^2}(1 + \sigma(\chi) -|\varphi|^2)^2 \bigg].
\end{split}
\end{equation}	
The corresponding self dual equations, which minimize $E_1$ to the value of $2\pi n$, are
\begin{subequations}\label{BPSCS}
		\begin{align}
		&B=\frac{2}{\kappa^2}|\varphi|^2(1 + \sigma -|\vfi|^2),\\ 
		&(D_{1} + iD_2)\varphi=0, 
		\end{align}	
\end{subequations}
which must be solved together with~\eqref{FOchi}. Additionally, the fields must be subject to the Chern-Simons constraint $\kappa B=2A_{0}|\varphi|^2$. The ensuing static equations give rise to Chern-Simons vortices in the presence of impurities, such as those considered in Ref.~\cite{Han}. In the interest of brevity, we shall not present solutions of~\eqref{BPSCS} for specific choices of $\sigma(\chi)$, but we note that the extension of the analysis we have developed throughout this work to solutions of~\eqref{BPSCS} is straightforward.

In this paper, we have investigated a model where a neutral scalar field is coupled to a Maxwell-Higgs subsystem and shown that, in the Bogomol'nyi limit, the equations of motion become those of the impurity model investigated in Refs.~\cite{Hook,Tong}. When the self-dual equations~\eqref{BPS} are solved, the scalar field may be found by integration of a first order differential equation, and the result substituted in $\sigma(\chi)$, thus giving rise to a function of the coordinates, which may be interpreted as an impurity if $\chi$ is a static defect that is stable under small perturbations. To circumvent Derrick's theorem and find a static configuration, we have used a $\widetilde{V}$ of the form introduced in~\cite{stable}, which allows for a stable, radially symmetric kink-like defect that solves a first order equation in the radial coordinate. These results have been illustrated through a family of impurities specified by Eq.~\eqref{sigma}. In the examples considered, the Lagrangian is symmetric under the change $\chi\to-\chi$, thus giving rise to theory invariant under $\rm{U(1)}\times Z_2$ transformations. Rotationally symmetric solutions for these models have been found numerically, and it was revealed that many properties induced by the impurity on the $\rm{U(1)}$ subsystem may be understood in relation to internal parameters of the kink. Finally, we showed how this method may be extended to Chern-Simmons-Higgs vortices.

Perspectives include the extension of these results to symmetries other than $\rm{U(1)}\times Z_2$, in which case the vortices may be coupled to other types of defects. One important alternative is found when one exchanges $E_2$ by an energy functional of the Maxwell-Higgs/Ginzburg-Landau type. In that case, $\chi$ must be taken as a complex field minimally coupled to a gauge field $\mathcal{A_{\mu}}$, while $\sigma$ may be chosen as a function of $|\chi|$, in order to preserve invariance of the Lagrangian under $\rm{U(1)}\times \rm{U(1)}$ transformations. In that case, the procedure developed in this paper remains essentially unchanged, as one may still minimize $E_2$ independently of $E_1$, and substitute the result into~\eqref{BPS}, which will, in that case, be coupled to the well-known self-dual equations found in Maxwell-Higgs theory~\cite{bogo}. This possibility may have important applications, as local $\rm{U(1)}\times \rm{U(1)}$ gauge invariance has been considered in investigations of dark matter~\cite{Witten, Long,DM,SchapoI, SchapoII, ahep}. For a $\sigma$ function of the form~\eqref{delta} (with $\chi$ exchanged by $|\chi|$), such a model would allow for a very interesting impurity of the form  $\sigma(\mathbf{x})=2\pi m\sum_{\mathbf{\tilde{x}_0}} \delta(\mathbf{x}-\mathbf{\tilde{x}_0})$, corresponding to multiple delta impurity sources located at the points $\tilde{x}_0$, which specify the distinct zeros of $\chi$.

Our procedure leads to a radially symmetric impurity, but a more general dependence on the coordinates could in principle be achieved by means of different choices of $\widetilde{V}$. An investigation of the dynamics of the models proposed in this work is also an interesting perspective. In particular, when the kink is allowed to move, the impurity is expected to change, as illustrated by the dependence of $\sigma$ on the position of the zero of $\chi$. Thus, vortex scattering is expected to be different from that of Tong and Wong's work when time-dependence is allowed for the neutral scalar field. Finally, we note that the generalization of this procedure to a curved space may bring interesting consequences. Vortices in a Riemannian manifold under the effect of impurities have been investigated in~\cite{Cockburn}. In our method, the equation of motion of the neutral scalar field would be changed in the presence of curvature, so that different geometries are expected to lead to different solutions $\chi(\mathbf{x})$, thus changing the resulting impurity.

\flushleft\textsl{\Huge}{\textbf{Declaration of competing interest}}

\vspace{5pt}

The authors declare that they have no known competing financial interests or personal relationships that could have appeared to influence the work reported in this paper.

\acknowledgements{Work supported by the Brazilian agencies Coordena\c{c}\~ao de Aperfei\c{c}oamento de Pessoal de N\'ivel Superior (CAPES), grant No 88887.485504/2020-00 (MAL), Conselho Nacional de Desenvolvimento Cient\'ifico e Tecnol\'ogico (CNPq), grants No. 404913/2018-0 (DB) and No. 303469/2019-6 (DB), Federal University of Para\'\i ba (UFPB/PROPESQ/PRPG) project code PII13363-2020 and Paraiba State Research Foundation (FAPESQ-PB) grant No. 0015/2019.}

%\biem{tdift} Reeves B. D., \textit{An introduction to topological defects in field theories}
%\bibitem{HDR2009_4} Halliday, David; Resnick, Robert, Walker, Jearl. Fundamentos de Física. 8ª ed. LTC, 2009. Vol. 04.

%\bibitem{Eisberg}R. Eisberg, R. Resnick, Física Quântica, Ed. Campus, Rio de Janeiro, 1979.
%\bibitem{Zettili}Zettili, Nouredine (2009). Quantum Mechanics: Concepts and Applications (2nd ed.). Chichester: Wiley. p. 365
%\bibitem{Lo

\end{document}